\documentclass[aip, amsmath,amssymb, reprint]{revtex4-1}

\usepackage{graphicx}
\usepackage{dcolumn}
\usepackage{bm}
\usepackage[utf8]{inputenc}
\usepackage[T1]{fontenc}
\usepackage{mathptmx}
\usepackage{etoolbox}
\usepackage{xcolor}
\usepackage{tabularx,booktabs}
\usepackage{amsfonts}
\usepackage{amsmath}
\usepackage{gensymb}
\usepackage{hyperref}

\begin{document}

\title{Physics-Integrated Inference for Signal Recovery in Non-Gaussian Regimes}

\author{Mohamed A. Mousa}
\author{Leif Bauer}
\author{Ziyi Yang}
\author{Utkarsh Singh}
\author{Angshuman Deka}
\author{Zubin Jacob}
\email{zjacob@purdue.edu}
\affiliation{Elmore Family School of Electrical and Computer Engineering and Birck Nanotechnology Center, Purdue University, West Lafayette, IN 47907, USA}

\date{\today}

\begin{abstract}
High-performance room-temperature sensing is often limited by non-stationary $1/f$ fluctuations and non-Gaussian stochasticity. In spintronic devices, thermally activated Néel switching creates heavy-tailed noise that masks weak signals, defeating linear filters optimized for Gaussian statistics. Here, we introduce a physics-integrated inference framework that decouples signal morphology from stochastic transients using a hierarchical 1D CNN-GRU topology. By learning the temporal signatures of Néel relaxation, this architecture reduces the Noise Equivalent Differential Temperature (NEDT) of spintronic Poisson bolometers by a factor of six (233.78 mK to 40.44 mK), effectively elevating room-temperature sensitivity toward cryogenic limits. We demonstrate the framework's universality across the electromagnetic and biological spectrum, achieving a 9-fold error suppression in Radar tracking, a 40\% uncertainty reduction in LiDAR, and a 15.56 dB SNR enhancement in ECG. This hardware-inference coupling recovers deterministic signals from fluctuation-dominated regimes, enabling near-ideal detection limits in noisy edge environments.
\end{abstract}

\maketitle

The sensitivity of room-temperature sensing is fundamentally constrained by stochastic thermodynamics. In high-performance regimes, the detection limit is often set not by the fundamental Johnson-Nyquist noise, but by excess non-stationary $1/f$ fluctuations and non-Gaussian transients \cite{das2023thermodynamically,le2018impact,bianconi2020recent}. In emerging spintronic devices, thermally activated Néel-type magnetic switching creates discrete, heavy-tailed noise that masks weak signals well above the theoretical thermal floor \cite{rogalski2019infrared}. Standard linear filters, which are statistically optimized for additive Gaussian noise, fail to reject these discrete transients without severely distorting the signal bandwidth. Consequently, approaching the fundamental thermodynamic limit ($k_B T$) has historically necessitated cryogenic cooling to physically suppress these fluctuations, restricting field deployability \cite{kanai2021theory,yang2025optical}.

Standard methodologies for sensor stabilization rely on heuristic estimators, such as $\sigma$-clipping \cite{ozdougru2025automatic}, median filtering \cite{guan2024adaptive}, or adaptive Kalman-based variants \cite{qi2025adaptive,liu2025deepgins}. Although computationally efficient, these methods operate on the premise of local signal stationarity, modeling the raw transduction $y(t)$ as:
\begin{equation}
\hat{s}(t) = H[y(t)] = H[s(t) + \eta(t) + \xi(t)]
\end{equation}
where $s(t)$ is the physical state, $\eta(t)$ is additive Gaussian thermal noise, and $\xi(t)$ denotes non-Gaussian transient noise. Critically, because the operator $H$ relies on local windowing or Gaussian error minimization, it lacks the temporal memory required to distinguish the physical transient $\xi(t)$ from high-frequency signal components. This imposes a fundamental trade-off where outlier suppression occurs at the expense of bandwidth and phase-latency \cite{yun2025information}.

In spintronic Poisson bolometers \cite{bauer2025exploiting,mousa2026ultra,yang2025long,singh2025long}, the noise component $\xi(t)$ is intrinsic to thermally activated magnetization switching, governed by the N\'eel relaxation law \cite{kanai2021theory}. Conventional filters lack the capacity to model these underlying dynamics, preventing standard improvements in Poissonian sensors. Consequently, while tailored metasurfaces have successfully integrated spectral identification directly into hardware \cite{mousa2021toward}, hardware design alone cannot eliminate intrinsic switching noise. Overcoming this barrier therefore requires a transition from static physical filtering to dynamic, physics-integrated inference \cite{karniadakis2021physics, wetzstein2020inference, barbastathis2019use}.

\begin{figure}[htbp]
    \centering
    \includegraphics[width=0.4\textwidth]{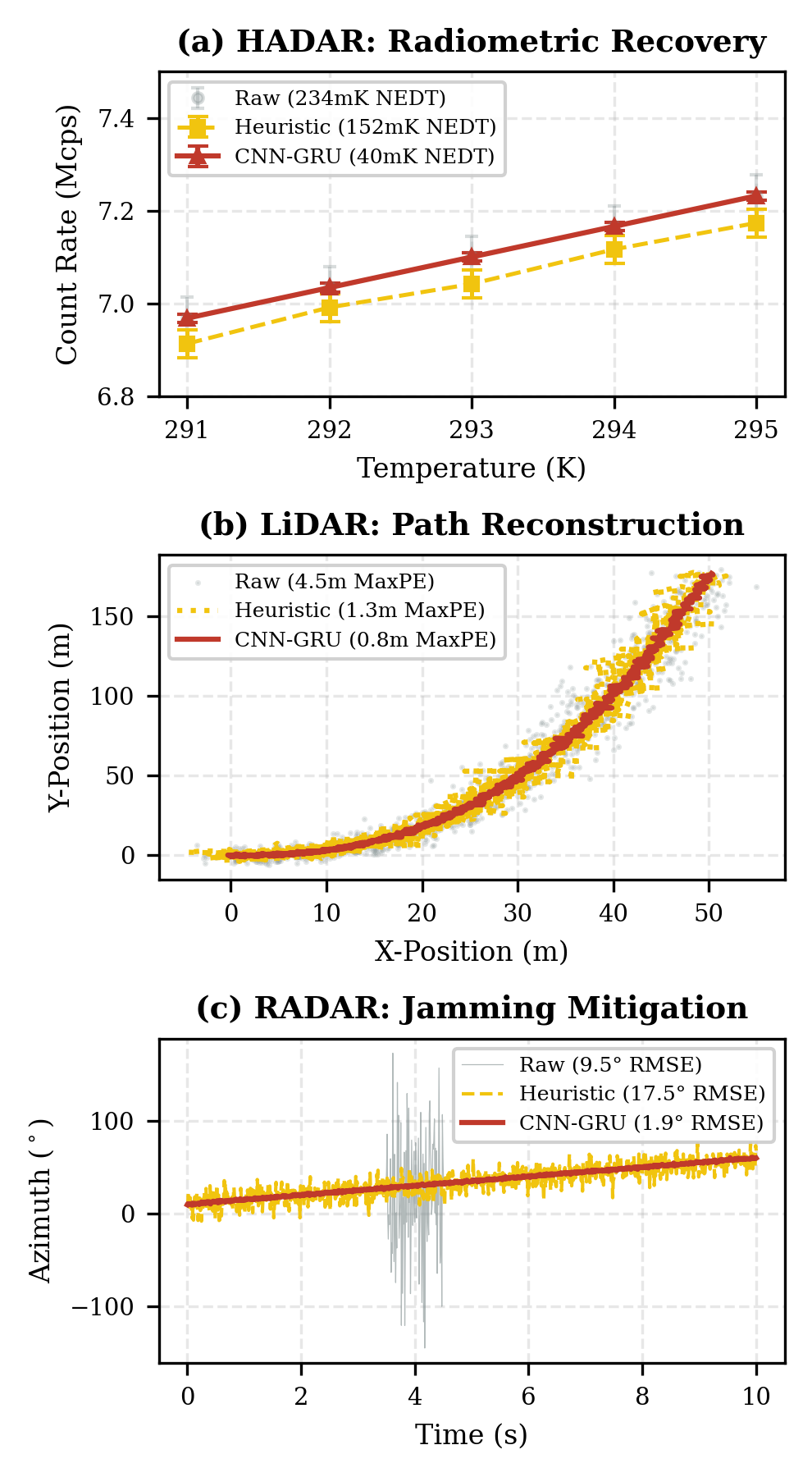}
    \caption{General signal restoration. Physics-integrated inference (red) resolves deterministic signals from stochastic inputs (grey), outperforming heuristic baselines (yellow). (a) Radiometric recovery in HADAR, reducing NEDT from 234 mK to 40 mK and avoiding heuristic signal attenuation. (b) LiDAR path reconstruction reducing MaxPE to 0.8 m. (c) Radar jamming mitigation maintaining 1.9$^\circ$ RMSE stability against electronic bursts.}
    \label{fig:Vision}
\end{figure}

In this work, we present a general, physics-integrated architecture designed to map raw hardware transduction directly to estimates approaching fundamental thermodynamic limits (Fig. \ref{fig:Vision}). Our framework utilizes a hierarchical 1D Convolutional Neural Network (CNN) and Bidirectional Gated Recurrent Unit (Bi-GRU) topology. This design is specifically engineered to exploit the time-domain dynamics and stochastic statistics of physical sensors \cite{raissi2019physics}, integrating short-term feature extraction with long-term history. Unlike causal heuristic filters that introduce phase-lag \cite{widmann2015digital}, this architecture resolves non-Gaussian transitions in real-time by leveraging bidirectional context \cite{wang2022novel}. We validate this hardware-inference co-design by demonstrating a nearly six-fold reduction in the Noise Equivalent Differential Temperature (NEDT) of spintronic sensors, suppressing the noise floor from 233.78 mK to 40.44 mK. This result demonstrates that room temperature sensor performance can be elevated toward the cryogenic sensing regime by effectively removing excess stochastic noise.

We demonstrate the universality of this framework by applying it to diverse physical domains beyond thermal sensing. In ranging applications, including Heat-Assisted Detection and Ranging (HADAR) \cite{bao2023heat}, LiDAR, and Radar, the architecture resolves spatial targets obscured by environmental interference and electronic jamming. Similarly, in biomedical monitoring, it isolates cardiac signal structures from high-amplitude motion artifacts in electrocardiography (ECG) (Table \ref{tab:results}). In contrast to non-linear adaptive filters that often induce baseline shifts or amplitude distortion \cite{widmann2015digital, li2021automatic}, our approach preserves radiometric linearity. This ensures that the reconstructed output maintains a strict proportional relationship to the physical input, allowing the sensor to retain its calibration and quantitative validity even in fluctuation-dominated environments.

The spintronic Poisson bolometer serves as a definitive model system for validating physics-integrated inference due to its stochastic-dominated transduction mechanism \cite{bauer2025exploiting}. In contrast to conventional infrared sensors that utilize the linear bolometric modulation of electrical resistance \cite{rogalski2025room}, the spintronic bolometer operates via thermally activated magnetization switching between two discrete stable states, $M_1$ and $M_2$. Incident radiation induces a localized temperature increment $\Delta T$ that reduces the Arrhenius-type energy barrier $E_b$ separating these magnetic orientations. This thermal modulation increases the instantaneous switching probability, encoding the radiometric flux into a sequence of stochastic transitions governed by Néel-Brown dynamics. Because the signal of interest is embedded in the temporal statistics of these discrete events—rather than a continuous voltage potential—this system represents an ideal physical candidate for demonstrating the reconstruction of deterministic signals from non-Gaussian noise regimes.

\begin{figure}[htbp]
    \centering
    \includegraphics[width=0.4\textwidth]{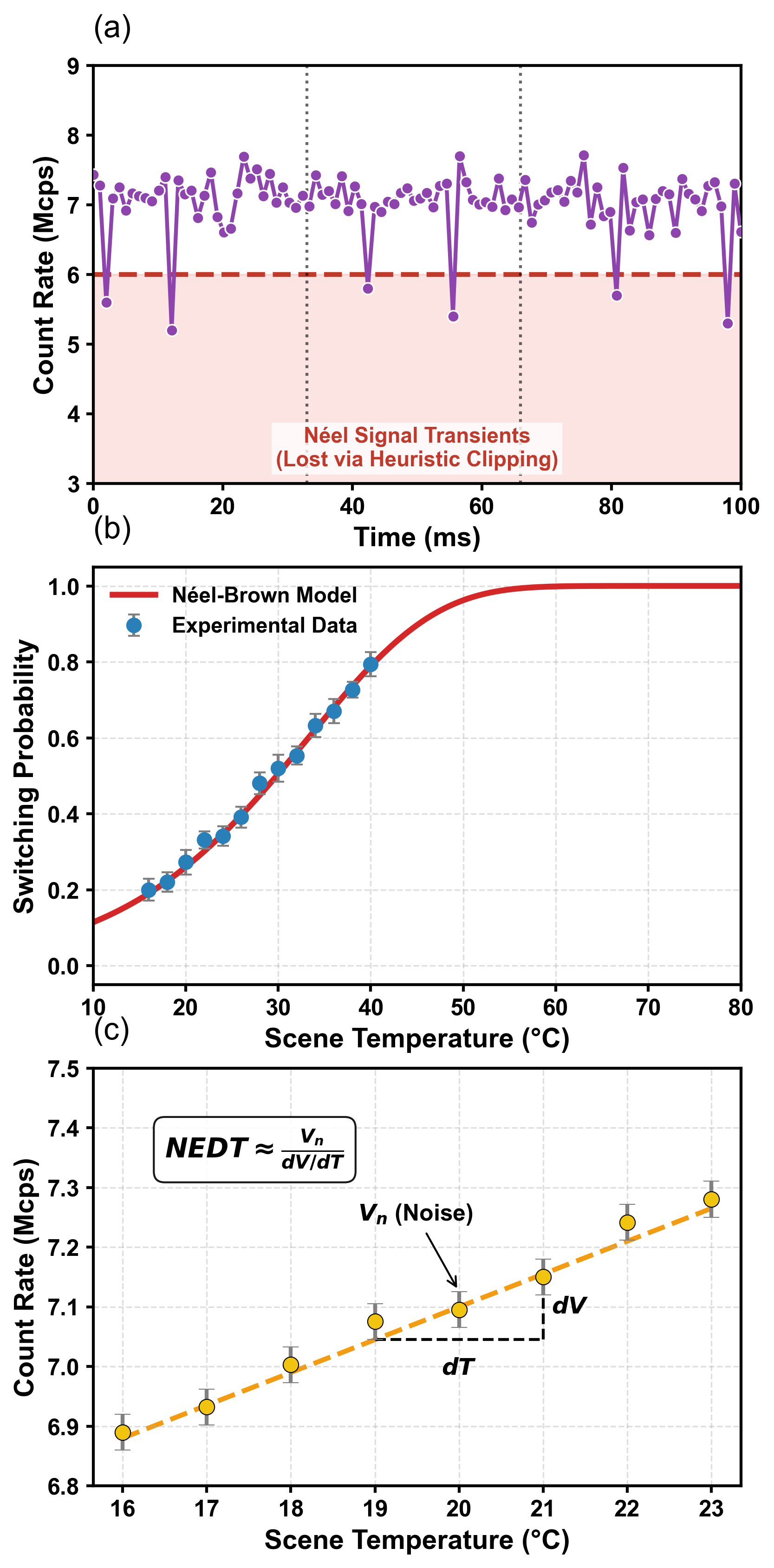}
    \caption{Stochastic physics of spintronic transduction. (a) Time-domain dynamics: Raw transients containing low-amplitude Néel signal events (red zone) that are misclassified and lost via standard heuristic clipping. (b) Thermally activated reversal: Switching probability follows an exponential Arrhenius signature (red curve), confirming the signal's physical origin. (c) Radiometric transfer function: The Noise Equivalent Differential Temperature (NEDT) is defined by the ratio of the RMS noise floor ($V_n$) to the linear sensor responsivity ($dV/dT$, dashed line).}
    \label{fig:Outlier_Physics}
\end{figure}

This mechanism is mathematically described by the Néel-Brown relaxation law \cite{coffey2012langevin}, where the relaxation time $\tau$ follows an Arrhenius dependence:
\begin{equation}
\tau(T) = \tau_0 \exp\left(\frac{E_b(T)}{k_B T}\right)
\end{equation}
where $\tau_0$ is the attempt period, $k_B$ is the Boltzmann constant, and $T$ is the absolute temperature. In the temporal domain, this physical process manifests as Random Telegraph Noise (RTN). Since the magnetization reversal is nearly instantaneous relative to the thermal integration time, these discrete state jumps appear as low-amplitude transients that punctuate the baseline signal (Fig. \ref{fig:Outlier_Physics}a).

Crucially, the root-mean-square (RMS) noise voltage $V_n$ in this regime is dominated by these discrete switching events rather than background Johnson noise. The switching probability $P_{sw}$ within an observation interval $\Delta t$ is given by:
\begin{equation}
P_{sw}(T) = 1 - \exp\left(-\frac{\Delta t}{\tau(T)}\right)
\end{equation}
As illustrated in Fig. \ref{fig:Outlier_Physics}b, the statistical distribution of these switching events relative to the scene temperature exhibits a characteristic exponential Arrhenius signature, validating the model of thermally activated reversal under constant magnetic anisotropy \cite{brown1963thermal}. This non-linear probability profile establishes the physical baseline for non-heuristic recovery.

The primary metric for thermal sensitivity, the Noise Equivalent Differential Temperature (NEDT), is defined as the ratio of the noise voltage $V_n$ to the sensor responsivity $\mathcal{R}$:
\begin{equation}
NEDT = \frac{V_n}{\mathcal{R}} = \frac{V_n}{\left| \partial V / \partial T \right|}
\end{equation}
where $\partial V / \partial T$ is the slope of the linear radiometric transfer function (Fig. \ref{fig:Outlier_Physics}c). Heuristic filters typically fail in this regime because they treat the stochastic transitions comprising $V_n$ as external outliers to be clipped, rather than state-dependent signals. Consequently, while adaptive thresholding may provide superficial noise reduction, it fundamentally compromises the radiometric transfer function by discarding valid Néel signal transients. This leads to signal attenuation and a failure to reach theoretical detection limits \cite{vollmer2018infrared}, necessitating the proposed hierarchical architecture to resolve the underlying dynamics of Néel-driven transitions.

\begin{figure}[htbp]
    \centering
     \includegraphics[width=0.48\textwidth]{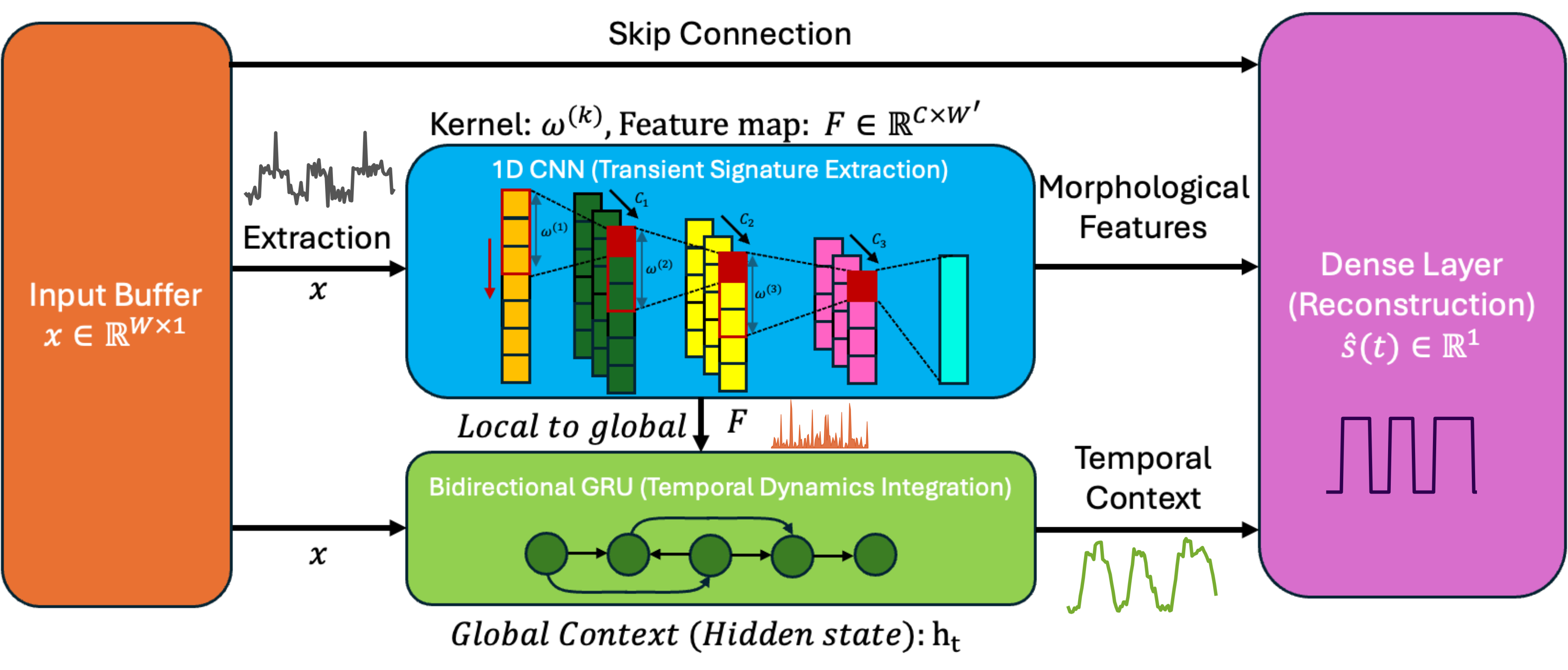}
    \caption{Hierarchical Spatiotemporal Reasoning Core. (Top) Skip connection preserves signal integrity. (Center) 1D-CNN extracts transient signatures into feature map $F \in \mathbb{R}^{C \times W'}$ (where $C$ is channel depth, $W'$ is encoded time). (Bottom) Bi-GRU utilizes global hidden state $h_t$ to resolve history-dependent stochasticity. A dense layer fuses these streams to recover physical estimate $\hat{s}(t)$ from input $\mathbf{x} \in \mathbb{R}^{W \times 1}$.}
    \label{fig:Architecture}
\end{figure}

To address the stochastic limitations defined by the Néel-Brown relaxation physics, we developed a hierarchical inference framework designed to maximize signal fidelity in uncooled regimes (Fig. \ref{fig:Architecture}). Moving beyond scalar amplitude thresholding, which indiscriminately clips low-amplitude data, this architecture exploits the distinct temporal statistics of the transduction noise \cite{borders2019integer, sidi2026tunable}. The framework maps the raw, non-Gaussian transduction $y(t)$ to a deterministic physical estimate $\hat{s}(t)$ through three specialized computational stages.

First, to perform \textit{local feature extraction (CNN)}, the primary stage utilizes a 1D Convolutional Neural Network (CNN) to isolate the characteristic dynamics of magnetization switching. For an input buffer $\mathbf{x} \in \mathbb{R}^{W \times 1}$, the feature maps are generated via convolution with learnable kernels optimized to recognize the specific step-like morphology of Random Telegraph Noise (RTN). Unlike fixed-weight heuristic filters, these kernels adapt to the macroscopic manifestation of discrete Néel reversal events \cite{chen2025stochastic, lambert2020random}, effectively decoupling these physical transients from uncorrelated electronic background noise \cite{bai2025high}.

Second, to enable \textit{global temporal reasoning}, we employ a Bidirectional Gated Recurrent Unit (Bi-GRU). The GRU maintains a hidden state $\mathbf{h}_t$ that serves as a temporal memory of the thermal context. By processing the sequence in both forward and backward temporal directions, the architecture acquires a global context of the signal trajectory \cite{lin2024bi}, enabling the model to resolve ambiguous transients by evaluating signal stability across the entire observation window \cite{nga2025hybrid}.

Finally, for \textit{high-fidelity reconstruction}, a dense projection layer is used to map the abstract spatiotemporal features back into the continuous physical domain. As illustrated in Fig. \ref{fig:Architecture}, we implement a skip connection that preserves raw signal integrity, ensuring the reconstruction layer retains access to the original signal scale. This hierarchical fusion allows for the recovery of radiometric linearity and minimizes estimation variance, specifically in regimes where traditional heuristic filters cause signal attenuation (Fig. \ref{fig:Vision}a).

\begin{figure}[htbp]
    \centering
    \includegraphics[width=0.4\textwidth]{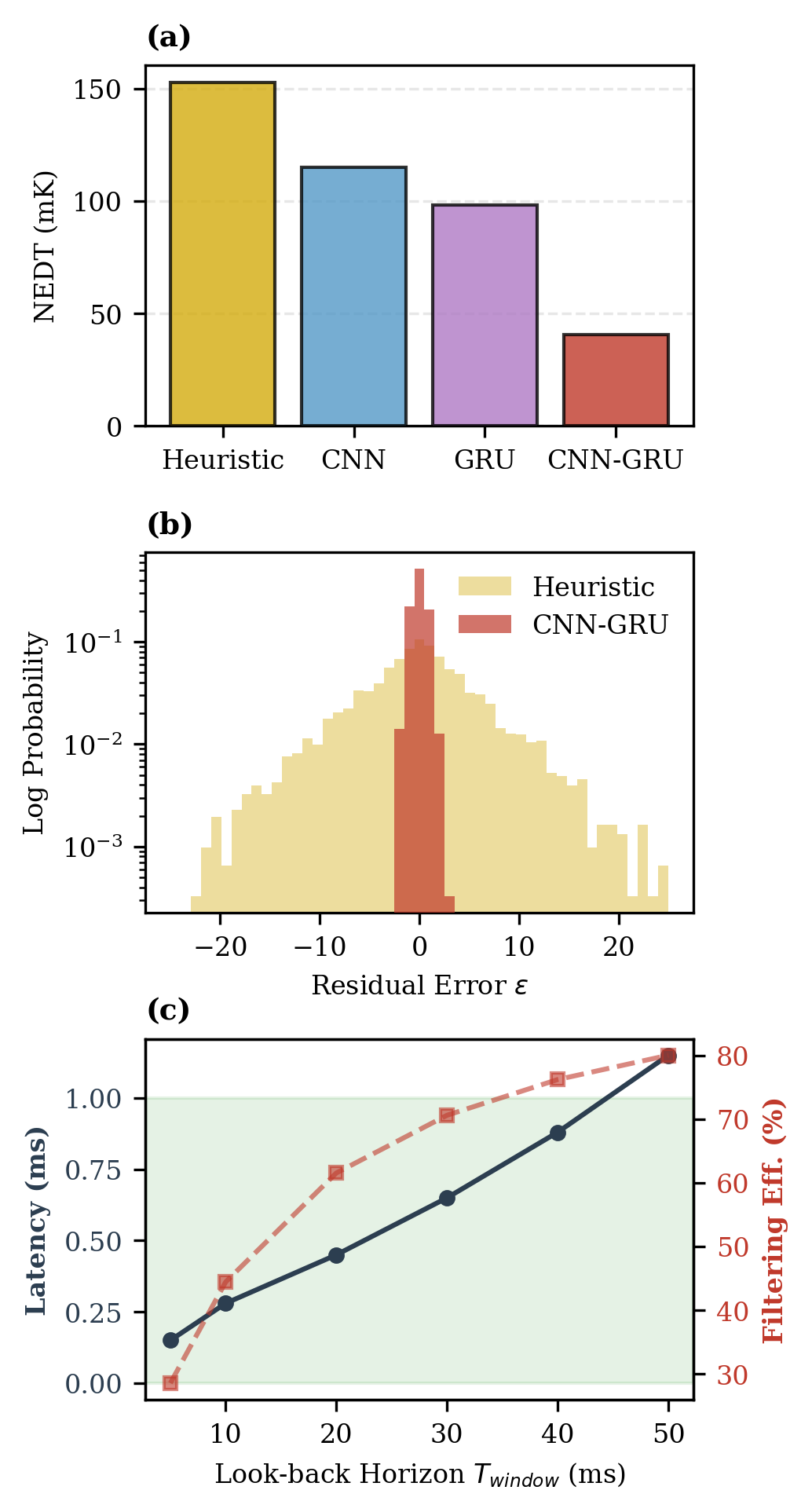}
    \caption{Performance sensitivity. (a) Ablation study confirming the CNN-GRU hybrid is required for <45 mK noise. (b) Residual error density $\epsilon$; inference core (red) Gaussianizes the heavy-tailed transients seen in heuristic filtering (yellow). (c) Latency vs. efficiency; green zone marks real-time operation ($L < 1$ ms).}
    \label{fig:Efficiency}
\end{figure}

To evaluate the efficacy of the physics-integrated inference core, we performed a comparative analysis against state-of-the-art heuristic filters across four distinct physical modalities. The primary benchmark, involving the spintronic Poisson bolometer, demonstrates the framework’s capacity to restore fundamental sensor performance. As illustrated in the thermal transfer function (Fig. \ref{fig:Vision}a), raw transduction suffers from significant count-rate suppression and non-linearity at elevated temperatures due to stochastic saturation. While heuristic adaptive algorithms provide marginal stabilization, they fail to recover the linear responsivity $\mathcal{R} = |\partial V / \partial T|$ of the device. In contrast, the CNN-GRU framework restores radiometric linearity, resulting in a six-fold reduction in the Noise Equivalent Differential Temperature (NEDT). As summarized in Table \ref{tab:results}, the NEDT is reduced from a raw value of 233.78 mK to 40.44 mK, nearly bridging the performance gap between room-temperature spintronics and cryogenically cooled standards.

The robustness of the architecture extends beyond thermal sensing to diverse time-series and spatial data streams, validating its general utility. In cardiac monitoring, where high-amplitude motion artifacts typically induce signal saturation, the proposed framework successfully isolates the underlying cardiac morphology to achieve a Signal-to-Noise Ratio (SNR) of 15.56 dB—a substantial improvement over the -3.24 dB observed with heuristic methods. This performance gain is mirrored in navigational point clouds, where the CNN-GRU architecture identifies environmental outliers as non-physical transients, reducing the Maximum Position Error (MaxPE) in LiDAR path reconstruction from 4.46 m to 0.75 m (Fig. \ref{fig:Vision}b). Furthermore, under conditions of extreme electronic jamming in Radar azimuth tracking, the bidirectional GRU utilizes global temporal context to maintain tracking through high-power bursts (Fig. \ref{fig:Vision}c), restoring a coherent signal path with a Root Mean Square Error (RMSE) of only 1.92$^\circ$.

We further performed an architecture sensitivity analysis (Fig. \ref{fig:Efficiency}a) to isolate the contribution of specific computational blocks to the noise suppression mechanism. The results indicate that the hybrid CNN-GRU configuration is strictly required to minimize the thermal noise floor below 45 mK. Isolated architectures fail to distinguish signal from fluctuation: the CNN alone interprets high-frequency thermal fluctuations as valid structural texture, failing to reject rapid transients. Conversely, the GRU integration provides the necessary temporal context, utilizing the history of the thermal state to identify and reject these fluctuations as stochastically uncorrelated noise. This hardware-inference synergy is confirmed by the residual error distribution $\epsilon$ (Fig. \ref{fig:Efficiency}b), where the inference core effectively Gaussianizes the heavy-tailed Néel noise profile. Finally, latency analysis (Fig. \ref{fig:Efficiency}c) validates the system for real-time edge deployment, maintaining sub-millisecond inference speeds ($L < 1$ ms) across the operational look-back horizon \cite{shi2016edge}.

\begin{table}[htbp]
\caption{\label{tab:results}Performance comparison across physical modalities. Values in bold indicate the superior performance of the physics-driven CNN-GRU framework.}
\begin{ruledtabular}
\begin{tabular}{lcccc}
\textbf{Case} & \textbf{Metric} & \textbf{Raw} & \textbf{Heuristic} & \textbf{CNN-GRU} \\
\hline
Thermal & NEDT (mK) & 233.78 & 152.82 & \textbf{40.44} \\
ECG & SNR (dB) & 9.75 & -3.24 & \textbf{15.56} \\
LiDAR & MaxPE (m) & 4.46 & 1.26 & \textbf{0.75} \\
Radar & RMSE ($^\circ$) & 9.45 & 17.49 & \textbf{1.92} \\
\end{tabular}
\end{ruledtabular}
\end{table}

Our results demonstrate that intrinsic Néel-driven switching noise is not a fundamental hardware constraint but a resolvable temporal signature. By partitioning noise mitigation into the inference domain, we transform this stochastic bottleneck into a deterministic reconstruction task that preserves radiometric linearity, avoiding the artifacts inherent to heuristic filters. This approach enables uncooled, CMOS-compatible neural bolometers \cite{mousa2025neural} to approach cryogenic sensitivities, achieving a six-fold reduction in NEDT (from 233.78 mK to 40.44 mK) without bulky cooling infrastructure. Validated across fluctuation-dominated regimes, yielding significant gains in ECG (15.56 dB SNR), LiDAR (0.75 m error), and Radar (1.92$^\circ$ RMSE). This hardware-inference co-design establishes a pathway to reaching theoretical detection limits in noisy edge environments.

\section*{Acknowledgment}
This work was partially supported by an Elmore Chaired Professorship at Purdue University.

\section*{AUTHOR DECLARATIONS}
\subsection*{Conflict of Interest}
The authors have no conflicts to disclose.

\section*{References}

\bibliography{aipsamp}

@book{rogalski2019infrared,
  title={Infrared and terahertz detectors},
  author={Rogalski, Antoni},
  year={2019},
  publisher={CRC press}
}

@article{kanai2021theory,
  title={Theory of relaxation time of stochastic nanomagnets},
  author={Kanai, Shun and Hayakawa, Keisuke and Ohno, Hideo and Fukami, Shunsuke},
  journal={Physical Review B},
  volume={103},
  number={9},
  pages={094423},
  year={2021},
  publisher={APS}
}

@article{bao2023heat,
  title={Heat-assisted detection and ranging},
  author={Bao, Fanglin and Wang, Xueji and Sureshbabu, Shree Hari and Sreekumar, Gautam and Yang, Liping and Aggarwal, Vaneet and Boddeti, Vishnu N and Jacob, Zubin},
  journal={Nature},
  volume={619},
  number={7971},
  pages={743--748},
  year={2023},
  publisher={Nature Publishing Group UK London}
}

@article{yang2025optical,
  title={Optical Response in Spintronic Poisson Bolometers},
  author={Yang, Ziyi and Gupta, Sakshi and Shalabi, Jehan and Bauer, Leif and He, Daien and Mousa, Mohamed and Deka, Angshuman and Jacob, Zubin},
  journal={arXiv preprint arXiv:2512.14968},
  year={2025}
}

@article{bauer2025exploiting,
  title={Exploiting Spintronics at Room Temperature for Long-Wave Infrared Nanophotonic Digital Bolometers},
  author={Bauer, Leif and Deka, Angshuman and Mousa, Mohamed A and Gupta, Sakshi and He, Daien and Huang, Sijay and Prasad, Bhagwati and Santos, Tiffany and Ray, Biswajit and Jacob, Zubin},
  journal={Nano Letters},
  volume={25},
  number={14},
  pages={5599--5608},
  year={2025},
  publisher={ACS Publications}
}

@article{raissi2019physics,
  title={Physics-informed neural networks: A deep learning framework for solving forward and inverse problems involving nonlinear partial differential equations},
  author={Raissi, Maziar and Perdikaris, Paris and Karniadakis, George E},
  journal={Journal of Computational physics},
  volume={378},
  pages={686--707},
  year={2019},
  publisher={Elsevier}
}

@book{coffey2012langevin,
  title={The Langevin equation: with applications to stochastic problems in physics, chemistry and electrical engineering},
  author={Coffey, William and Kalmykov, Yu P},
  volume={27},
  year={2012},
  publisher={World Scientific}
}

@book{vollmer2018infrared,
  title={Infrared thermal imaging: fundamentals, research and applications},
  author={Vollmer, Michael and M{\"o}llmann, Klaus-Peter},
  year={2018},
  publisher={John Wiley \& Sons}
}

@article{mousa2025neural,
  title={Neural bolometers: Designing next-generation infrared thermal imagers with in-pixel neuromorphic computing},
  author={Mousa, Mohamed A and Singh, Utkarsh and Bauer, Leif and Deka, Angshuman and Jacob, Zubin},
  journal={Physical Review Applied},
  volume={24},
  number={6},
  pages={064044},
  year={2025},
  publisher={APS}
}

@article{mousa2021toward,
  title={Toward spectrometerless instant Raman identification with tailored metasurfaces-powered guided-mode resonances (GMR) filters},
  author={Mousa, Mohamed A and Rafat, Nadia H and Saleh, Amr AE},
  journal={Nanophotonics},
  volume={10},
  number={18},
  pages={4567--4577},
  year={2021},
  publisher={De Gruyter}
}

@article{liu2025deepgins,
  title={DeepGINS: A Deep Learning-Based Approach to Robust Virtual INS/GNSS Positioning in Non-Gaussian Noise},
  author={Liu, Yuxuan and Hu, Xuguang and Zhang, Juwei and Liu, Bo and Ren, Bingyi},
  journal={IEEE Internet of Things Journal},
  year={2025},
  publisher={IEEE}
}

@article{yun2025information,
  title={Information Perception Adaptive Filtering Algorithm Sensitive to Signal Statistics: Theory and Design},
  author={Yun, Shiwei and Guan, Sihai and Zhao, Yong and Xiang, Qiang and Zhang, Chuanwu and Biswal, Bharat},
  journal={Mathematics},
  volume={13},
  number={20},
  pages={3294},
  year={2025},
  publisher={MDPI}
}

@book{rogalski2025room,
  title={Room Temperature Photon Detectors},
  author={Rogalski, Antoni and Hu, Weida and Martyniuk, Piotr},
  year={2025},
  publisher={CRC Press}
}

@article{le2018impact,
  title={Impact of fundamental thermodynamic fluctuations on light propagating in photonic waveguides made of amorphous materials},
  author={Le Thomas, Nicolas and Dhakal, Ashim and Raza, Ali and Peyskens, Fr{\'e}d{\'e}ric and Baets, Roel},
  journal={Optica},
  volume={5},
  number={4},
  pages={328--336},
  year={2018},
  publisher={Optical Society of America}
}

@article{bianconi2020recent,
  title={Recent advances in infrared imagers: toward thermodynamic and quantum limits of photon sensitivity},
  author={Bianconi, Simone and Mohseni, Hooman},
  journal={Reports on Progress in Physics},
  volume={83},
  number={4},
  pages={044101},
  year={2020},
  publisher={IOP Publishing}
}

@article{das2023thermodynamically,
  title={Thermodynamically limited uncooled infrared detector using an ultra-low mass perforated subwavelength absorber},
  author={Das, Avijit and Mah, Merlin L and Hunt, John and Talghader, Joseph J},
  journal={Optica},
  volume={10},
  number={8},
  pages={1018--1028},
  year={2023},
  publisher={Optica Publishing Group}
}

@article{karniadakis2021physics,
  title={Physics-informed machine learning},
  author={Karniadakis, George Em and Kevrekidis, Ioannis G and Lu, Lu and Perdikaris, Paris and Wang, Sifan and Yang, Liu},
  journal={Nature Reviews Physics},
  volume={3},
  number={6},
  pages={422--440},
  year={2021},
  publisher={Nature Publishing Group UK London}
}

@article{wetzstein2020inference,
  title={Inference in artificial intelligence with deep optics and photonics},
  author={Wetzstein, Gordon and Ozcan, Aydogan and Gigan, Sylvain and Fan, Shanhui and Englund, Dirk and Solja{\v{c}}i{\'c}, Marin and Denz, Cornelia and Miller, David AB and Psaltis, Demetri},
  journal={Nature},
  volume={588},
  number={7836},
  pages={39--47},
  year={2020},
  publisher={Nature Publishing Group UK London}
}

@article{barbastathis2019use,
  title={On the use of deep learning for computational imaging},
  author={Barbastathis, George and Ozcan, Aydogan and Situ, Guohai},
  journal={Optica},
  volume={6},
  number={8},
  pages={921--943},
  year={2019},
  publisher={Optical Society of America}
}

@inproceedings{ozdougru2025automatic,
  title={Automatic detection and characterization of random telegraph noise in sCMOS sensors},
  author={{\"O}zdo{\u{g}}ru, Arda and Karpov, Sergey and Christov, Asen and V{\'\i}tek, Stanislav},
  booktitle={Optical Sensors 2025},
  volume={13527},
  pages={191--202},
  year={2025},
  organization={SPIE}
}

@article{guan2024adaptive,
  title={Adaptive median filter salt and pepper noise suppression approach for common path coherent dispersion spectrometer},
  author={Guan, Shouxin and Liu, Bin and Chen, Shasha and Wu, Yinhua and Wang, Feicheng and Liu, Xuebin and Wei, Ruyi},
  journal={Scientific Reports},
  volume={14},
  number={1},
  pages={17445},
  year={2024},
  publisher={Nature Publishing Group UK London}
}

@article{qi2025adaptive,
  title={Adaptive Kalman Filters based on Elliptically Contoured Distributions for Heavy-Tailed and Non-Stationary Measurement Noise},
  author={Qi, Bin and Zhang, Songyuan and Chen, Weihan and Fu, Yili and Ren, Bingyin},
  journal={IEEE Transactions on Instrumentation and Measurement},
  year={2025},
  publisher={IEEE}
}

@article{widmann2015digital,
  title={Digital filter design for electrophysiological data--a practical approach},
  author={Widmann, Andreas and Schr{\"o}ger, Erich and Maess, Burkhard},
  journal={Journal of neuroscience methods},
  volume={250},
  pages={34--46},
  year={2015},
  publisher={Elsevier}
}

@article{wang2022novel,
  title={A novel end-to-end network based on a bidirectional GRU and a self-attention mechanism for denoising of electroencephalography signals},
  author={Wang, Wenlong and Li, Baojiang and Wang, Haiyan},
  journal={Neuroscience},
  volume={505},
  pages={10--20},
  year={2022},
  publisher={Elsevier}
}

@article{li2021automatic,
  title={An automatic method to reduce baseline wander and motion artifacts on ambulatory electrocardiogram signals},
  author={Li, Hongzu and Boulanger, Pierre},
  journal={Sensors},
  volume={21},
  number={24},
  pages={8169},
  year={2021},
  publisher={MDPI}
}

@article{brown1963thermal,
  title={Thermal fluctuations of a single-domain particle},
  author={Brown Jr, William Fuller},
  journal={Physical review},
  volume={130},
  number={5},
  pages={1677},
  year={1963},
  publisher={APS}
}

@article{borders2019integer,
  title={Integer factorization using stochastic magnetic tunnel junctions},
  author={Borders, William A and Pervaiz, Ahmed Z and Fukami, Shunsuke and Camsari, Kerem Y and Ohno, Hideo and Datta, Supriyo},
  journal={Nature},
  volume={573},
  number={7774},
  pages={390--393},
  year={2019},
  publisher={Nature Publishing Group UK London}
}

@article{sidi2026tunable,
  title={Tunable random telegraph noise in stable perpendicular magnetic tunnel junctions for unconventional computing},
  author={Sidi El Valli, Ahmed and Tsao, Michael and Chen, Dairong and Kent, Andrew D},
  journal={Physical Review Applied},
  volume={25},
  number={1},
  pages={014035},
  year={2026},
  publisher={APS}
}

@article{chen2025stochastic,
  title={Stochastic Nature of Voltage-Controlled Charge Dynamics in AlO x Magnetic Tunnel Junctions},
  author={Chen, Chun-Yen and Huang, Bao-Huei and Tang, Yu-Hui and Gonzalez-Ruano, C{\'e}sar and Aliev, Farkhad G and Ling, Dah-Chin and Hong, Jhen-Yong},
  journal={Nano Letters},
  year={2025},
  publisher={ACS Publications}
}

@article{lambert2020random,
  title={Random telegraph signal analysis with a recurrent neural network},
  author={Lambert, NJ and Esmail, AA and Edwards, M and Ferguson, AJ and Schwefel, HGL},
  journal={Physical Review E},
  volume={102},
  number={1},
  pages={012312},
  year={2020},
  publisher={APS}
}

@article{bai2025high,
  title={A High-Performance Training-Free Pipeline for Robust Random Telegraph Signal Characterization via Adaptive Wavelet-Based Denoising and Bayesian Digitization Methods},
  author={Bai, Tonghe and Kapoor, Ayush and Kim, Na Young},
  journal={arXiv preprint arXiv:2510.10752},
  year={2025}
}

@article{lin2024bi,
  title={A Bi-GRU-attention neural network to identify motor units from high-density surface electromyographic signals in real time},
  author={Lin, Chuang and Chen, Chen and Cui, Ziwei and Zhu, Xiujuan},
  journal={Frontiers in Neuroscience},
  volume={18},
  pages={1306054},
  year={2024},
  publisher={Frontiers Media SA}
}

@article{nga2025hybrid,
  title={Hybrid One-Dimensional Convolutional Neural Network—Recurrent Neural Network Model for Reconstructing Missing Data in Structural Health Monitoring Systems},
  author={Nga, Nguyen Thi Thu and Matos, Jose C and Ngoc, Son Dang},
  journal={Machines},
  volume={13},
  number={12},
  pages={1101},
  year={2025},
  publisher={MDPI}
}

@article{shi2016edge,
  title={Edge computing: Vision and challenges},
  author={Shi, Weisong and Cao, Jie and Zhang, Quan and Li, Youhuizi and Xu, Lanyu},
  journal={IEEE internet of things journal},
  volume={3},
  number={5},
  pages={637--646},
  year={2016},
  publisher={IEEE}
}

@article{yang2025long,
  title={Long-Wave Infrared Spintronic Poisson Bolometers with High Sensitivity},
  author={Yang, Ziyi and Gupta, Sakshi and Shalabi, Jehan and He, Daien and Bauer, Leif and Deka, Angshuman and Jacob, Zubin},
  journal={arXiv preprint arXiv:2512.12490},
  year={2025}
}

@article{mousa2026ultra,
  title={Ultra-broadband Mid to Long-wave Infrared Spintronic Poisson Bolometer},
  author={Mousa, Mohamed A and Bauer, Leif and He, Daien and Gupta, Sakshi and Jape, Shubhankar and Singh, Utkarsh and Prasad, Bhagwati and Mukherjee, Partha P and Deka, Angshuman and Jacob, Zubin},
  journal={arXiv preprint arXiv:2601.11733},
  year={2026}
}

@article{singh2025long,
  title={Long wave infrared detection using probabilistic spintronic bolometer arrays},
  author={Singh, Utkarsh and Bauer, Leif and Deka, Angshuman and Mousa, Mohamed and He, Daien and Gupta, Sakshi and Prasad, Bhagwati and Jacob, Zubin},
  journal={arXiv preprint arXiv:2510.06519},
  year={2025}
}

\end{document}